
\input eplain

\def\CurrentFontFile{}
\newdimen\DefaultFontSize
\DefaultFontSize=10pt
\newdimen\CurrentFontSize
\def\SetFont#1{\SetFontSpec{#1}{\CurrentFontSize}}

\def\SetFontSpec#1#2{\edef\CurrentFontFile{#1}\CurrentFontSize=#2\font\ActiveFont#1 at\CurrentFontSize\ActiveFont}
\SetFontSpec{ptmr}{\DefaultFontSize}

\def\rm{\fam0\SetFont{\BaseFont}}
\def\it{\fam\itfam\SetFont{\ItalicFont}}
\def\sl{\fam\slfam\SetFont{\SlantedFont}}
\def\bf{\fam\bffam\SetFont{\BoldFont}}

\def\SetTextFontGroup#1#2{\def\InstallProc{\csname InstallFontGroup#1\endcsname}\expandafter
\ifx\InstallProc\relax\errmessage{Unable to install font group #1}\else
\InstallProc{#2}\fi}

\def\SetMathFontGroup#1#2{\def\InstallProc{\csname InstallMathFontGroup#1\endcsname}\expandafter
\ifx\InstallProc\relax\errmessage{Unable to install math font group #1}\else
\InstallProc{#2}\fi}

\def\InstallMathFontGroupTimesOld#1{%
\font\tf=cmr12
\font\sf=cmr9
\font\ssf=cmr7
\textfont0=\tf 
\scriptfont0=\sf
\scriptscriptfont0=\ssf
\font\tf=cmmi12
\font\sf=cmmi9
\font\ssf=cmmi7
\skewchar\tf='177
\skewchar\sf='177
\skewchar\ssf='177
\textfont1=\tf 
\scriptfont1=\sf
\scriptscriptfont1=\ssf
\font\tfs=cmsy10 at 12pt
\font\sfs=cmsy9
\font\ssfs=cmsy7
\skewchar\tfs='60
\skewchar\sfs='60
\skewchar\ssfs='60
\textfont2=\tfs 
\scriptfont2=\sfs
\scriptscriptfont2=\ssfs
\textfont3=\tenex
\scriptfont3=\tenex
\scriptscriptfont3=\tenex
\font\tfsa=txsya at 12pt
\font\sfsa=txsya at 9pt
\font\ssfsa=txsya at 7pt
\textfont\msafam=\tfsa
\scriptfont\msafam=\sfsa
\scriptscriptfont\msafam=\ssfsa
\font\tfsb=txsyb at 12pt
\font\sfsb=txsyb at 9pt
\font\ssfsb=txsyb at 7pt
\textfont\msbfam=\tfsb
\scriptfont\msbfam=\sfsb
\scriptscriptfont\msbfam=\ssfsb}

\newdimen\mtsize
\newdimen\mssize
\newdimen\msssize
\newfam\msafam
\newfam\msbfam

\def\Bbb#1{{\fam\msbfam\relax#1}}

\SetTextFontGroup{Times}{\DefaultFontSize}
\SetMathFontGroup{Times}{\DefaultFontSize}

\input amssym.tex
\DefaultFontSize=12pt
\SetTextFontGroup{Times}{12pt}
\SetMathFontGroup{Times}{12pt}

\def\InstallTimesMath#1{\SetMathFontGroup{Times}{#1}}

\vsize=8.6in
\hsize=6in
\hoffset 0.25 in
\voffset 0.25 in
\parskip = 8 pt
\parindent = 0pt

\newcount\sectioncount
\newcount\subsectioncount

\def\seclabel#1{\definexref{#1}{\number\sectioncount}{sec}}

\def\bl{}
\def\el{}
\def\li#1{\par\penalty500\hangindent=4em
\hangafter=0\indent\llap{\SetFontSpec{ptmr}{12pt}#1\hskip 0.5em}}
\def\define#1{#1}%

\newif\ifdraft
\drafttrue
\def\Version#1{\draftfalse\gdef\versiondate{#1}}

\newif\iflatexing\latexingfalse
\newif\ifjournal\journalfalse

\def\section#1{\advance\sectioncount by 1 \subsectioncount = 0 \proccount=0
\vskip 10pt plus 6pt minus 2pt \goodbreak
{\InstallTimesMath{16pt}\SetFontSpec{pplb}{16pt}\number\sectioncount. #1\par}}

\def\subsection#1{\advance\subsectioncount by 1
\vskip 5pt plus 3pt \goodbreak
{\InstallTimesMath{13pt}\SetFontSpec{pplb}{13pt}
\number\sectioncount.\number\subsectioncount\ #1\par}}

\def\startmatter{\bgroup\noindent\vskip 2cm plus 0.5cm 
\baselineskip=12pt
\ifx\titletext{}\else{\raggedright\SetFontSpec{pplb}{20pt}\hbox to \hsize{\vbox{\titletext}}}\bigskip\fi
\SetFontSpec{ptmr}{12pt}%
\InstallTimesMath{12pt}%
\ifx\authortext{}\else{\bf \authortext}\hfill
\vskip 0pt{\authorhome\hfill}
\smallskip\fi
\ifdraft Draft: \today\else\versiondate\fi\vskip 1cm
\ifx\abstracttext{}\else{{\SetFontSpec{pplb}{10pt}Abstract. 
}\SetFontSpec{ptmr}{10pt}\abstracttext\smallskip}\fi
\egroup\advance \baselineskip by 4pt}
\let\endmatter\bye

\def\title#1{\global\def\titletext{#1}}
\def\titletext{}
\def\Author#1#2#3{\global\def\authortext{#1}\gdef\authorhome{#2}\gdef\authoremail{#3}}
\def\authortext{}
\def\authorhome{}
\def\Abstract#1{\global\def\abstracttext{#1}}
\def\abstracttext{}
\def\headtitle{{\SetFontSpec{ptmri}{10pt}\titletext\ifdraft\ (Draft: \today)\fi}} 
\def\headauthor{{\SetFontSpec{ptmri}{10pt}\authortext}} 

\headline={\ifnum\pageno>1\SetFontSpec{ptmr}{10pt}%
\ifx\titletext{}\folio.\hfill\else
\ifodd\pageno\headtitle\hfill\folio\else
\folio\hfill\headauthor\fi\fi\else\hfil\fi}
\nopagenumbers

\catcode`@=11 
\def\makeheadline{\vbox to\z@{\vskip-35\p@
  \line{\vbox to8.5\p@{}\the\headline}\vss}\nointerlineskip}
\catcode`@=12 

\catcode`@=11 
\def\providecommand#1{
   \def\@commandname{#1}%
   \@getoptionalarg\@postprovidecommand
}
\def\@postprovidecommand#1{}
\catcode`@=12 
\def\textbf#1{{\bf#1}}

\newcount\proccount
\def\myproclaim#1{\advance\proccount by 1\medbreak\noindent{\bf #1.%
\enspace}\bgroup\sl}
\def\procnumber{\number\sectioncount.\number\proccount}

\def\endmyproclaim{\egroup\par
\ifdim\lastskip<\smallskipamount \removelastskip\penalty55\smallskip\fi}

\def\Proof{\endmyproclaim{\it Proof:\hskip 0.5em}}
\def\NoProof{\endmyproclaim}
\def\EOProof{\unskip\quad$\square$\goodbreak\par}

\def\propword{Proposition}
\def\proplabel#1{\definexref{#1}{\procnumber}{prop}}
\def\prop{\myproclaim{\propword\ \procnumber}\def\label{\proplabel}}

\def\theoremword{Theorem}
\def\theoremlabel#1{\definexref{#1}{\procnumber}{theorem}}
\def\thm{\myproclaim{\theoremword\ \procnumber}\def\label{\theoremlabel}}

\def\lemmaword{Lemma}
\def\lemmalabel#1{\definexref{#1}{\procnumber}{lemma}}
\def\lem{\myproclaim{\lemmaword\ \procnumber}\def\label{\lemmalabel}}

\def\corword{Corollary}
\def\corlabel#1{\definexref{#1}{\procnumber}{cor}}
\def\cor{\myproclaim{\corword\ \procnumber}\def\label{\corlabel}}

\def\Remark{{\bf Remark.}\enspace}
\def\EORemark{\smallskip\par}

\def\({$$\eqalign\begingroup}
\def\){\endgroup$$}

\def\Acknowledge{\par{\bf Acknowledgements.}\enspace}
\def\EOAcknowledge{\par}

\catcode`@=11 

\def\eqaligntd#1{\null\,\vcenter\bgroup\openup\jot\m@th
  \ialign\bgroup\strut\hfil$\displaystyle{##}$&$\displaystyle{{}##}$&$\displaystyle{{}##}$\hfil
\crcr}
\def\endeqaligntd{\crcr\egroup\egroup\,}

\def\eqalignd{\null\,\vcenter\bgroup\openup\jot\m@th
  \ialign\bgroup\strut\hfil$\displaystyle{##}$&$\displaystyle{{}##}$\hfil
      \crcr}      
\def\endeqalignd{\crcr\egroup\egroup\,}

\def\eqalignnod{\displ@y \tabskip\centering
  \halign to\displaywidth\bgroup\hfil$\@lign\displaystyle{##}$\tabskip\z@skip
    &$\@lign\displaystyle{{}##}$\hfil\tabskip\centering
    &\llap{$\@lign##$}\tabskip\z@skip\crcr}
\def\endeqalignnod{\crcr\egroup}

\catcode`@=12 

\def\be{$$}
\def\ee{$$}
\def\bel#1{\gdef\eqlabel{#1}$$}
\def\eel{\eqdef{\eqlabel}$$}
\def\bea{\bgroup\let\\\cr$$\eqalignd}
\def\eea{\endeqalignd$$\egroup}
\def\beal#1{\bgroup\let\\\cr\gdef\eqlabel{#1}$$\eqalignd}
\def\eeal{\endeqalignd\eqdef{\eqlabel}$$\egroup}
\def\beat{\bgroup\let\\\cr$$\eqaligntd}
\def\eeat{\endeqaligntd$$\egroup}
\def\beatl#1{\bgroup\let\\\cr\gdef\eqlabel{#1}$$\eqaligntd}
\def\eeatl{\endeqaligntd\eqdef{\eqlabel}$$\egroup}
\def\bean{\bgroup\let\\\crcr\let\label\ealabel
$$\eqalignnod}
\def\eean{\endeqalignnod$$\egroup}
\def\ealabel#1{&\eqdef{#1}\crcr}

\let\ref\refn
\def\nonumber{}


\def\sf{\SetFontSpec{phvr}{10pt}}

\def\ds{\displaystyle}
\def\newpage{\vfill\eject}
\def\til{\lower 0.75ex\hbox{\char126}}

\def\W#1#2{W^{{#1},{#2}}}


\def\Lap{\Delta\,}

\def\emph#1{{\it#1}}

\def\references{\newpage\section{References}}

\input epsf

\def\epsboxw#1#2{{\epsfxsize=#1\epsfbox{#2}}}
\newcount\figurecount
\def\figlabel#1{\definexref{#1}{\number\figurecount}{fig}}
\def\fig#1#2{\advance\figurecount by 1\noindent\vskip\parskip\bigskip
{\def\label{\figlabel}\vbox{\hbox to \hsize{\hfil#1\hfil}\par\hfil
Figure \number\figurecount: #2\hfil\bigskip}}}
\long\def\omit#1\endomit{}

\def\gflat{{\overline g} }
\def\Reals{{\Bbb R}}
\def\Rn{\Reals^n}
\def\abs#1{\left|#1\right|}
\def\partialbeta{\partial^{\,\beta}}
\def\loc{_{\rm loc}}

\def\cpct{_{\rm c}}

\def\gchi{\raise 0.4 ex\hbox{$\chi$}}
\def\evat#1#2{\left.#1\;\right|_{#2}}

\def\calP{{\cal P}}
\def\calM{{\cal M}}

\def\del{_\delta}

\def\srho{_{\rho}}

\def\tr{\mathop{\rm tr}}
\def\div{\mathop{\rm div}}
\def\ondm{\quad\hbox{on $\partial M$}}
\def\tg{{\tilde{g}}}
\def\tR{{\tilde{R}}}

\def\tu{\tilde{u}}
\def\hatg{{\hat g}}
\def\hatK{{\hat K}}

\def\ck{{\Bbb L}}
\def\vl{\Delta_{\Bbb L}}

\def\Leps{\vl}
\def\Beps{B\,}
\def\Lflat{\overline{\Delta}_{\Bbb L}}
\def\Bflat{\overline{B}}
\def\vlflat{\overline{\Delta}_{\Bbb L}}
\def\ckflat{\overline{\Bbb L}}
\def\jia{_j^{\,i,\alpha}}

\def\jiapa{\jia\partial_\alpha}

\def\hBr{B_r^+}
\def\dV{\;dV}
\def\dA{\;dA}

\def\converges#1{\mathrel{\ds\mathop{\longrightarrow}_{#1}}}

\def\PL{{\cal P}_{\Bbb L}}
\def\Lie#1{{\cal L}_{#1}}
\def\dimc{\kappa}
\def\expfrac#1#2{{#1\over#2}}
\def\ip<#1,#2>{\left<#1,#2\right>}
\def\P#1#2#3{{\calP}^{#1,#2}_{#3}}

\Version{June 22, 2004}
\title{Solutions of the Einstein Constraint Equations \break
with Apparent Horizon Boundaries}
\ifjournal
\Author{David Maxwell}{Department of Mathematics, University of Washington,  
Seattle, Washington, 98195-4350, USA}{dmaxwell@math.washington.edu}
\else
\Author{David Maxwell}{University of Washington}{}
\fi
\Abstract{We construct asymptotically Euclidean solutions of the 
vacuum Einstein constraint equations with an apparent horizon boundary 
condition.  Specifically, we give sufficient conditions for the 
constant mean curvature conformal method to generate such solutions. 
The method of proof is based on the barrier method 
used by Isenberg for compact manifolds without boundary, suitably
extended to accommodate semilinear boundary conditions and low 
regularity metrics.  As a consequence of our results for manifolds
with boundary, we also obtain improvements to the theory of the constraint 
equations on asymptotically Euclidean manifolds without boundary.
}
\startmatter

\section{Introduction}
The $N$-body problem in general relativity concerns the 
dynamics of an isolated system of $N$ black holes.   
One aspect of the problem, quite different from its classical counterpart, 
is the complexity of constructing appropriate initial data for the associated
Cauchy problem.  Initial data on an $n$-manifold $M$ is a 
Riemannian metric $g$ and a symmetric $(0,2)$-tensor $K$.  
We think of $M$ as an embedded spacelike hypersurface of an ambient 
Lorentz manifold ${\cal M}$; $g$ is the pullback of the the Lorentz 
metric on $\cal M$ and $K$ is the extrinsic curvature of $M$ in ${\cal M}$.
To model an  isolated system of $N$ black holes in vacuum, 
the triple $(M,g,K)$ must satisfy several requirements.  Isolated systems are 
typically modeled with asymptotically Euclidean initial data.  This requires 
that $g$ approach the Euclidean metric and that $K$ decay to zero at far 
distances in $M$  (see Section \ref{sec:ae} for a rigorous definition).  
Moreover, the vacuum Einstein equation imposes a compatibility
condition on $K$, $g$, and its scalar curvature $R$
\beal{constraints}
R-\abs{K}^2+\tr K ^2 &= 0\\
\div K -d\,\tr K &= 0.
\eeal
These are known as the Einstein constraint equations. Finally, data for 
$N$ black holes must evolve into a spacetime $\cal M$ containing an 
event horizon, and the intersection of the event horizon with $M$ 
must have $N$ connected components.  

An event horizon is the boundary of the region that can send signals 
to infinity.  It is a global property of a spacetime
and cannot be located in an initial data set without evolving the data.
This poses a serious obstacle to creating multiple black hole initial
data.  To address this problem, schemes such as those in \cite{misner-geostat},
\cite{bl-geostat}, \cite{bowen-york} and \cite{thornberg-ah} (see also
\cite{cook-bh}) create initial data containing an apparent horizon, defined
below.  The motivation for using apparent horizons comes from the weak
Cosmic Censorship conjecture.  Assuming that asymptotically Euclidean
initial data evolves into a weakly censored spacetime,  any apparent 
horizons present in the initial data will be contained in the black hole 
region of the spacetime.  To generate spacetimes with multiple black holes, 
one constructs initial data with multiple apparent horizons.  If the 
apparent horizons are well separated, one conjectures that they 
will be associated with distinct black holes.

An interesting approach to creating initial data containing apparent horizons,
first suggested by Thornburg \cite{thornberg-ah}, is to work with a 
manifold with boundary and prescribe that the boundary be an apparent 
horizon.  Thornburg numerically investigated 
generating such initial data.
Variations of the apparent horizon condition have subsequently
been proposed for numerical study, e.g. \cite{cook-binaries} 
\cite{eardley-bh}.  However, as indicated by Dain \cite{dain}, 
there has not been a rigorous mathematical investigation of the apparent
horizon boundary condition. The goal of this paper is to take
an initial step in addressing the problem of constructing asymptotically
Euclidean Cauchy data satisfying the apparent horizon boundary condition.
We exhibit sufficient conditions for generating a family of this data.

An \define{apparent horizon} is a surface that is instantaneously 
neither expanding nor contracting as it evolves under the flow of its 
outgoing (to infinity) orthogonal null geodesics. On the boundary
of $M$, the expansion under this flow is given by the so-called convergence
\bel{outconv}
\theta_+ =  - \tr K + K(\nu, \nu) -(n-1)h,
\eel
where $\nu$ is the exterior unit normal of $\partial M$ and
$h$ is the mean curvature of $\partial M$ computed with respect
to $-\nu$.  Hence $\partial M$ is an apparent horizon if $\theta_+=0$.
More generally, we say that $\partial M$ is outer marginally trapped
if $\theta_+\le0$. The corresponding convergence for the incoming 
(from infinity) family of null geodesics is
\bel{inconv}
\theta_- =  - \tr K + K(\nu, \nu) +(n-1)h.
\eel
A surface is marginally trapped if both $\theta_+\le0$ and
$\theta_-\le0$.

There is not a consistent definition of an apparent horizon in the
literature.  Other definitions of an apparent horizon include the boundary 
of a trapped region or an outermost marginally trapped surface.  
All these structures imply the 
existence of a black hole in  the spacetime.  We work with the definition 
$\theta_+=0$ since it is a local property and forms a natural boundary 
condition.   Hence we seek asymptotically Euclidean data $(M,g,K)$ satisfying 
\beal{const-sys}
R-\abs{K}^2+\tr K ^2 &= 0\\
\div K -d\,\tr K &= 0 \\
- \tr K + K(\nu, \nu) -(n-1)h &= 0 \ondm. \\
\eeal

The conformal method of Lichnerowicz  \cite{lich-conformal}, 
Choquet-Bruhat and York \cite{cb-york-held} provides a natural approach 
to the problem. For simplicity we work with its constant mean curvature (CMC) formulation, 
under which the constraint equations decouple.
The conformal method seeks a solution of the form
$$
(\hatg,\hatK)=(\phi^{4\over n-2}g,\phi^{-2}\sigma+{\tau\over n}g),
$$ 
where $g$ is a given asymptotically Euclidean metric prescribing the 
conformal class of $\hatg$, $\tau$ is a constant 
specifying $\tr \hatK$,  $\sigma$ is an unknown traceless symmetric
$(0,2)$-tensor, and $\phi$ is an unknown conformal factor tending to 1 at
infinity. From the decay conditions 
on $\hatK$ for asymptotically Euclidean initial data and the 
assumption that $\tau$ 
is constant, we have the further simplification $\tau=0$. 
Equations \eqref{const-sys} then 
become a semilinear equation with semilinear boundary condition for~$\phi$
\beal{conformal-h}
-\Lap \phi + {1\over a}\left(R\phi -\abs{\sigma}^2\phi^{-3-2\dimc}\right) &= 0\\
\partial_\nu\phi+{1\over\dimc} 
h\phi-{2\over a}\sigma(\nu,\nu)\phi^{-1-\dimc} &= 0
\quad\hbox{on $\partial M$}\\
\eeal
and a linear system for $\sigma$
\beal{conformal-m}
\div \sigma = 0.
\eeal
The first equation of \eqref{conformal-h} is known as the Lichnerowicz
equation. In \eqref{conformal-h}, 
the dimensional constants are  $\dimc=2/(n-2)$ and $a=2\dimc+4$.
Note that we use the exterior normal $\nu$ to follow 
traditional PDE notation, but the mean curvature $h$ is computed with respect 
to the interior unit normal.

A trace-free, symmetric $(0,2)$-tensor $\sigma$  satisfying \eqref{conformal-m}
is called \define{transverse traceless}.  The set of transverse traceless 
tensors forms a linear space, and the choice of $\sigma$ can be thought of as 
data to be prescribed in solving \eqref{conformal-h}.  Hence, we wish
to find conditions on $g$ and $\sigma$ under which \eqref{conformal-h} 
is solvable.

The construction of solutions of \eqref{conformal-h} starts with an
asymptotically Euclidean manifold $(M,g')$ satisfying
\bel{hyp1}
\lambda_{g'}>0.
\eel
For any asymptotically Euclidean metric $(M,g)$,  $\lambda_g$ is the 
conformal invariant
\be
\lambda_{g}=\inf_{f\in C\cpct^{\infty}(M), f\not\equiv0} {\int_M a \abs{\nabla f}^2 + R f^2\dV + \int_{\partial M} {a\over\dimc}h f^2\dA\over ||f 
||_{L^{2n/(n-2)}}^2}.
\ee
This is analogous to an invariant for compact manifolds with 
boundary found in \cite{escobar}. We next make a conformal change from $g'$
to a metric $g$ satisfying $R=0$ and $h<0$; Corollary~\ref{cor:guage2} ensures
this can always be done.  Our main result, Theorem~\ref{thm:main} proves
that \eqref{conformal-h} is then solvable if
\bel{hyp2}
(n-1)h\le\sigma(\nu,\nu)\le 0.
\eel
We prove Theorem \ref{thm:main} using a
barrier method for semilinear boundary conditions.  Section~\ref{sec:barrier}
establishes a general existence theorem and Section \ref{sec:lich} 
applies it to system~\eqref{conformal-h}.  

One can easily find asymptotically Euclidean manifolds with boundary 
that satisfy $\lambda_g>0$. For example, every manifold with $R\ge0$
and $h\ge0$ has $\lambda_g>0$.  So condition \eqref{hyp1} of the construction
can be readily met.  On the other hand, it is not obvious that the
restriction \eqref{hyp2} is reasonable.  In Section~\ref{sec:vl} we
show that on any sufficiently smooth asymptotically Euclidean manifold
with boundary we can freely specify $\sigma(\nu,\nu)$ on the boundary.
That is, given a function $f$ on $\partial M$, we can find a
transverse traceless tensor $\sigma$ satisfying $\sigma(\nu,\nu)=f$.  This 
follows from the solution to a boundary value problem for the vector Laplacian.
Since $h<0$, it follows that there exists a large family of transverse 
traceless tensors $\sigma$ satisfying \eqref{hyp2}.

It should be noted that our construction is not a full parameterization
of the set of CMC solutions of \eqref{const-sys}. Since condition 
\eqref{hyp2} is not conformally invariant, the set of allowed transverse 
traceless tensors depends on the choice of conformal representative 
satisfying $R=0$ and $h<0$.  Hence condition \eqref{hyp2} is not necessary.  
Moreover, although technical condition \eqref{hyp1} is vital for the 
construction, it is not clear if it is necessary.
Hence there remains much to be understood about parameterizing the
full set of solutions.

In light of recent low regularity a priori estimates
for solutions of the evolution problem
\cite{rough-einstein} \cite{smith-tataru-lwp}, there is interest in 
generating low regularity solutions of the  constraints.  In terms
of $L^p$ Sobolev spaces, a natural setting 
is $(g,\sigma)\in
\W 2{\expfrac{n}{2}+\epsilon}\loc\times\W1{\expfrac{n}{2}+\epsilon}\loc$.  
This is the weakest regularity that ensures that $g$ has curvature in
an $L^p$ space and that the Sobolev space containing
$g$ is an algebra.  Y.~Choquet-Bruhat has
announced \cite{cb-low-reg} a construction of such low regularity solutions of 
the constraint equations in the context of compact manifolds.
We construct asymptotically Euclidean solutions 
with this level of regularity,
but under a possibly unneeded assumption. 
In order to find suitable transverse traceless tensors, 
we require that $(M,g)$ not admit any nontrivial conformal Killing fields 
vanishing at infinity.  This is known to be true \cite{boost-problem} for $C^3$ 
asymptotically Euclidean manifolds.  We prove this also holds for metrics 
with regularity as weak  as $\W 2{n+\epsilon}$.  To consider 
$\W 2{\expfrac{n}{2}+\epsilon}$ metrics, however, we must assume the 
non-existence of such fields.

In a previous version of this article, we constructed solutions
of \eqref{const-sys} under the hypotheses $\lambda_g>0$ 
and  $\sigma(\nu,\nu)\ge0$.  
These solutions have the undesirable feature that although the boundary
is a outer marginally trapped surface, it is not a marginally trapped surface.
This observation was made in
\cite{dain-boundary}, which appeared shortly after our results were announced.
From \eqref{outconv} and \eqref{inconv} we see that
$\hat\theta_-=\hat\theta_+ +2(n-1)\hat h$.
Since $\hat\theta_+=0$ on $\partial M$, we have $\hat\theta_-=2(n-1)\hat h$.
So $\partial M$ is a marginally trapped surface if and only if
$\hat h \le 0$.  The sign of $\hat h$ is determined by $\sigma(\nu,\nu)$
since $\hat\theta_+=0$ implies
\be
(n-1)\hat h = \hat K(\hat \nu,\hat\nu) = \phi^{-2-2\dimc}\sigma(\nu,\nu).
\ee
So $\hat h\le0$ if and only if $\sigma(\nu,\nu)\le 0$.  Under the hypothesis
$\sigma(\nu,\nu)\ge0$, one can construct an apparent horizon that is also
a marginally trapped surface only if $\sigma(\nu,\nu)=0$, which
leads to $\hat\theta_-=0$ and $\hat h=0$.

In \cite{dain-boundary}, Dain constructs trapped surface boundaries by
working with $\hat\theta_-$ rather than $\hat\theta_+$.  Under suitable 
hypotheses, \cite{dain-boundary} prescribes 
$\hat\theta_-\le0$ and constructs boundaries with 
$\hat\theta_+\le\hat\theta_-\le 0$.  These are trapped surfaces,
but the inequality $\hat\theta_+\le\hat\theta_-$ shows that the resulting
boundaries satisfy $\hat h\ge 0$. In particular, the techniques of 
\cite{dain-boundary} also cannot construct a boundary 
that is simultaneously a marginally trapped surface and an apparent horizon 
unless $\hat\theta_-=0$ and $\hat h = 0$.

In the current version of this article, we construct solutions with 
$\hat h\le0$.   To do this requires we assume 
$\sigma(\nu,\nu)\le 0$, since the sign of $\hat h$ is determined by the
sign of $\sigma(\nu,\nu)$.  The resulting PDEs are more delicate,
and the hypothesis $(n-1)h\le\sigma(\nu,\nu)$ arises to compensate for this.
Since $\hat{h}\le 0$, the apparent horizons we construct satisfy
$\hat\theta_-\le\hat\theta_+=0$, with strict inequality wherever
$\sigma(\nu,\nu)$ (and hence $\hat h$) is negative.  Figure \ref{fig1} shows 
boundary mean curvatures of 
various signs and further indicates the naturality of the condition 
$\hat h \le 0$.

\fig{\epsboxw{4.5in}{hl.eps}}%
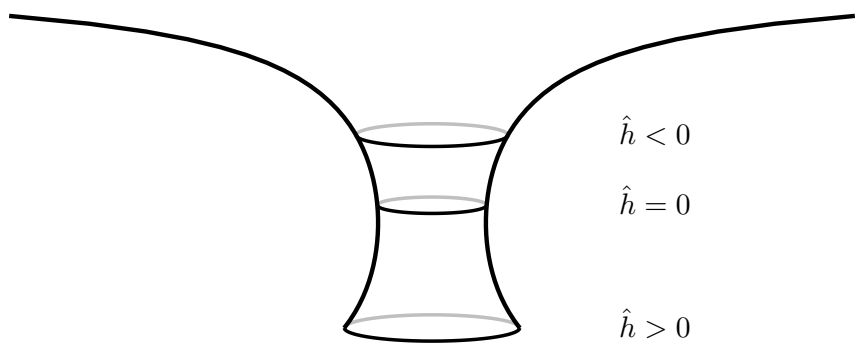
{Boundary mean curvatures of an asymptotically Euclidean manifold.\label{fig1}}

\subsection{Notation}
Let $\calM$ be a Lorentz manifold with metric $\gamma$ and connection $D$.  
The signature of $\gamma$ is $(-+\cdots+)$.
If $M$ is a spacelike hypersurface of $\calM$
with timelike unit normal $N$, we define the extrinsic curvature $K$ of
$M$ in $\calM$ by $K(X,Y)=\ip<D_X Y,N>_{\gamma}$ for vector fields tangent 
to $M$. This definition agrees with that used in \cite{bowen-york} and
\cite{dain}, but  differs in sign from that used in \cite{wald} and 
\cite{dain-boundary}.

Let $M$ be a Riemannian manifold with metric $g$ and connection $\nabla$,
If $\Sigma$ is a spacelike hypersurface if $M$ with unit normal $\nu$, 
the extrinsic curvature $k$ of $\Sigma$ in $M$ is similarly defined by 
$k(X,Y)=\ip<\nabla_X Y,\nu>_{g}$.  The mean curvature $h$ of $\Sigma$
computed with respect to $\nu$ is ${1\over n-1}\tr k$, where $n$
is the dimension of $M$.

Throughout this paper we take $n$ to be a fixed integer with $n\ge3$
and $\rho$ to be a negative real number.
If $p\in[1,\infty]$, we define the critical Sobolev exponent 
$p^*={np\over n-p}$ if $p<n$ and we set $p^*=\infty$ otherwise.
The ball of radius $r$ about $x$ is $B_r(x)$, and $E_r$ is the region
exterior to $\overline B_r(0)$.  We define $f^{(+)}(x)=\max(f(x),0)$.

We use the notation $A\lesssim B$ 
to mean $A< c B$ for a certain positive constant $c$.  The constant
is independent of the functions and parameters appearing in $A$
and $B$ that are not assumed to have a fixed value. For example, when
considering a sequence $\{f_i\}_{i=1}^\infty$ of functions on a 
domain $\Omega$, the expression
$||f_i||_{L^1(\Omega)}\lesssim 1$ means the sequence is uniformly 
bounded in $L^1(\Omega)$ (with a bound that might depend on $\Omega$).

\section{Asymptotically Euclidean Manifolds}
\seclabel{sec:ae}

An asymptotically Euclidean manifold is a non-compact Riemannian manifold,
possibly with boundary,  that can be decomposed into 
a compact core and a finite number of ends 
where the metric approaches the Euclidean metric at far distances.
To make this loose description precise, we use weighted
function spaces that prescribe asymptotic behavior like
$|x|^\delta$ for large $x$.  For $x\in\Rn$, let $w(x)=(1+|x|^2)^{1/2}$.
Then for any $\delta\in\Reals$ and and any open set $\Omega\subset\Rn$, the 
weighted Sobolev space $\W kp_\delta(\Omega)$ is the subset of 
$\W kp\loc(\Omega)$ for which the norm 
\bea
||u||_{\W kp_\delta(\Omega)} = 
\sum_{|\beta|\le k} ||w^{-\delta-\expfrac np+|\beta|}\,
\partialbeta u ||_{L^p(\Omega)}
\eea
is finite; we will always work with spaces for which $p\neq1,\infty$.
Weighted spaces of continuous functions are defined by the norm
\be
||u||_{C^k\del(\Omega)}=\sum_{\abs{\alpha}\le k} 
\sup_{x\in\Omega} w(x)^{-\delta+\abs{\alpha}}\abs{\partial^\alpha u(x)}.
\ee
Our indexing convention for $\delta$ follows \cite{mass-asympt-flat}
so that the value of $\delta$ directly encodes asymptotic growth at infinity.
We refer the reader to \cite{mass-asympt-flat} for properties of these
weighted spaces.  In particular, we recall the following facts.

\iflatexing\else\vskip -20pt\fi
\lem\label{basic}\iflatexing\strut\fi
\bl
\li{1.} If $p\le q$ and $\delta'<\delta$ then 
$L^p_{\delta'}(\Omega)\subset L^q\del(\Omega)$ and the inclusion is
continuous.
\li{2.} For $k\ge1$ and $\delta'<\delta$ the inclusion
$\W kp_{\delta'}(\Omega)\subset\W {k-1}p\del(\Omega)$ is compact.
\li{3.} If $1/p>k/n$ then $\W kp\del(\Omega)\subset L^r\del(\Omega)$
where $1/r=1/p-k/n$.  If $1/p=k/n$  then
$\W kp\del(\Omega)\subset L^r\del(\Omega)$ for all $r\ge p$.
If $1/p<k/n$ then $\W kp\del(\Omega)\subset C^0\del(\Omega)$.
These inclusions are continuous.
\li{4.} If $m\le\min(j,k)$, $p\le q$, $\epsilon>0$, and $1/q< 
(j+k-m)/n$, then
multiplication is a continuous bilinear map from
$\W jq_{\delta_1}(\Omega)\times\W kp_{\delta_2}(\Omega)$ to
$\W mp_{\delta_1+\delta_2+\epsilon}(\Omega)$.  In particular, 
if $1/p<k/n$ and $\delta<0$, then $\W kp\del(\Omega)$ is an algebra.

\el
\NoProof

Let $M$ be a smooth, connected, $n$-dimensional manifold
with boundary, and let $g$ be a metric on $M$ for which
$(M,g)$ is complete, and let $\rho<0$ (these will be standing 
assumptions for the remainder of the paper).  We say $(M,g)$ is 
\define{asymptotically Euclidean of class $\W kp\srho$} if:
\bl
\li{i.} The metric $g\in \W kp\loc(M)$, where $1/p-k/n<0$ (and 
consequently $g$ is continuous).
\li{ii.} There exists a finite collection $\{N_i\}_{i=1}^m$ of open subsets
of $M$ and diffeomorphisms $\Phi_i:E_1\mapsto N_i$ such that $M-\cup_i N_i$
is compact.
\li{iii.} For each $i$, $\Phi_i^*g-\gflat\in \W kp\srho(E_1)$.
\el

We call the charts $\Phi_i$ end charts and the corresponding coordinates
are end coordinates.  Suppose $(M,g)$ is asymptotically Euclidean, and 
let $\{\Phi_i\}_{i=1}^m$ be its collection of end charts.  
Let $K = M - \cup_i \Phi_i( E_2)$, so $K$
is a compact manifold with boundary.  The weighted Sobolev space
$\W kp_\delta(M)$ is the subset of $\W kp\loc(M)$ such that the
norm
\be
||u||_{\W kp_\delta(M)} = ||u||_{\W kp(K)} + \sum_i||\Phi_i^* u||_{\W 
kp\del(E_1)}
\ee
is finite.  The weighted spaces $L^p\del(M)$ and $C^k\del(M)$ are  defined 
similarly, and we let $C^\infty\del(M)=\cap_{k=0}^\infty C^k\del(M)$.
Lemma~\ref{basic} applies equally well to asymptotically Euclidean manifolds.

Using these weighted spaces we can now define an asymptotically
Euclidean data set.  The extrinsic curvature tensor $K$ of an initial 
data set $(M,g,K)$ should behave like a first derivative of $g$.  Hence, if
$(M,g)$ is asymptotically  Euclidean of class $\W kp\srho$, we say $(M,g,K)$
is an asymptotically Euclidean data set if $K\in\W{k-1}p_{\rho-1}(M)$.

\section{Barrier Method For Semilinear Boundary Conditions}
\seclabel{sec:barrier}
In \cite{const-compact}, Isenberg used a constructive barrier method 
(also known as the method of sub- and supersolutions) to completely parameterize
CMC solutions of the constraint equations on a compact manifold.  
Subsequently, the method has been applied to construct non-CMC solutions on
compact manifolds \cite{const-non-cmc-compact} and asymptotically
Euclidean solutions \cite{const-asympt-flat}.  In this section
we provide a version of the generic barrier construction that
accommodates semilinear boundary conditions and low regularity metrics.

Consider the boundary value problem
\beal{fund-sys}
-\Lap u & = F(x,u)\\
\partial_\nu u &=  f(x,u)\ondm
\eeal
on an asymptotically Euclidean manifold.  We use the convention
that $\Lap$ has negative eigenvalues, so 
$\Lap=\partial_{x_1}^2+\cdots+\partial_{x_n}^2$ in 
Euclidean space.  A \define{subsolution} of equation \eqref{fund-sys} 
is a function $u_-$ that satisfies
\bea
-\Lap u_- &\le  F(x,u_-)\\
\partial_\nu u_- &\le f(x,u_-)\ondm
\eea
and a \define{supersolution} $u_+$ is defined similarly with the inequalities 
reversed.  In Proposition \ref{prop:subsup} below, we show that if there exists a subsolution 
$u_-$ and a supersolution $u_+$ decaying at infinity and satisfying $u_-\le u_+$, 
then there exists a solution $u$ satisfying $u_-\le u \le u_+$. 

The proof of Proposition \ref{prop:subsup} relies on properties of the associated 
linearized operator
\beal{linearsys}
-\Delta u +V u & = F\\
\partial_\nu u+\mu u & = f,\ondm
\eeal
where $V$, $\mu$, $F$, and $f$ are functions of $x$ alone. 
Let $\calP$ denote
$\left((-\Delta+V),\evat{(\partial_\nu+\mu)}{\partial M}\right)$.  

\prop\label{prop:linearprob} Suppose $(M,g)$ is asymptotically 
Euclidean of class 
$\W kp\srho$, $k\ge2$, $k>n/p$, and 
suppose $V\in\W{k-2}{p}_{\rho-2}$ 
and $\mu\in\W{k-1-\expfrac1p}{p}$.  Then if 
$2-n<\delta<0$ the operator $\calP:\W {k}p\del(M)\rightarrow 
\W {k-2}p_{\delta-2}(M)\times\W {k-1-\expfrac1p}p(\partial M)$  is
Fredholm with index 0.
Moreover, if $V\ge0$ and $\mu\ge0$ then $\calP$ is an isomorphism.
\NoProof

In Proposition \ref{vliso} we prove a similar
result for the vector Laplacian.  Since the details are tedious and largely 
similar, we omit the proof of Proposition \ref{prop:linearprob}.  The only 
substantial 
difference from the proof of Proposition \ref{vliso} is the method used to
show $\calP$ 
is injective when $V$ and $\mu$ are nonnegative. This is an easy 
consequence of the following weak maximum principle.

\lem\label{p:maxpr} Suppose $(M,g)$, $V$, and $\mu$ satisfy the hypotheses
of Proposition \ref{prop:linearprob} and suppose $V\ge0$ and $\mu\ge0$.
If  $u\in\W kp\loc$ satisfies
\beal{maxprsys}
-\Delta u +V u & \le 0\\
\partial_\nu u+\mu u & \le 0,\ondm 
\eeal
and if $u^{(+)}$ is $o(1)$ on each end of $M$, then $u\le 0$.  
In particular, if $u\in\W kp\del(M)$
for some $\delta<0$ and $u$ satisfies \eqref{maxprsys}, then $u\le0$.
\Proof
Fix $\epsilon>0$, and let $v=(u-\epsilon)^{(+)}
$.  Since
$u^{(+)}=o(1)$ on each end, we see $v$ is compactly supported.
Moreover, since $u\in\W kp\loc$ we have from Sobolev embedding
that $u\in \W12\loc$ and hence $v\in\W12$. Now,
\be
\int_M -v\Delta u \dV \le -\int_M V u v\dV \le 0
\ee
since $V\ge0$, $v\ge 0$ and since $u$ is positive wherever $v\neq0$.
Integrating by parts we have
\be
\int_M |\nabla v|^2\dV -\int_{\partial M} v\partial_\nu u \dV \le 0,
\ee
since $\nabla u = \nabla v$ on the support of $v$.  From the boundary
condition we obtain
\be
\int_M |\nabla v|^2\dV \le -\int_{\partial M} \mu u  v\dV \le 0,
\ee
since $\mu\ge 0$.  So $v$ is constant and compactly supported, and we 
conclude $u\le\epsilon$.  Sending $\epsilon$ to 0 proves $u\le 0$.

Finally, if $u\in\W kp\del$, then $u\in C^0\del$.  Hence if $\delta<0$, then
$u^{(+)}=o(1)$ and the lemma can be applied to $u$.\EOProof

If $V$ or $\mu$ is negative at some point, the kernel of $\calP$
might not be empty. We have the following estimate for how elements
of the kernel decay at infinity.

\lem\label{kerneldecay} Suppose $(M,g)$, $V$, and $\mu$ satisfy
the hypotheses of Proposition \ref{prop:linearprob} and suppose that
$u\in\W kp\del$ with $\delta<0$ is in the kernel of $\calP$.
Then $u\in\W kp_{\delta'}(M)$  for every $\delta'\in(2-n,0)$.
\Proof
Since $V\in\W {k-2}p_{\rho-2}$ we have $Vu\in\W {k-2}p_{\rho+\delta-2}$.
Hence 
\bea
(-\Delta u,\partial_\nu u) & =(-Vu,-\mu u) \nonumber\\ 
&\in\W {k-2}p_{\rho+\delta-2}(M)\times\W{k-1-\expfrac1p}p(\partial M).\nonumber
\eea
Since $(-\Delta,\left(\evat{\partial_\nu}{\partial M}\right))$ is an 
isomorphism on $\W kp_{\delta'}$ for each $\delta'\in(2-n,0)$,
we conclude that $u\in\W kp_{\delta'}$ for each 
$\delta'\in(\max(2-n,\rho+\delta-2), 0)$.  Iterating this argument
a finite number of times yields the desired result.
\EOProof

Although the barrier construction in Proposition \ref{prop:subsup} below 
only uses the weak maximum principle
Lemma \ref{p:maxpr}, we need the following strong maximum principle 
in our later analysis of the Lichnerowicz equation to ensure that the 
conformal factors we construct never vanish. Note that there is no
sign restriction on $V$ and $\mu$.

\lem\label{p:strmaxpr} Suppose $(M,g)$, $V$, and $\mu$ satisfy
the hypotheses of Proposition \ref{prop:linearprob}.  Suppose also
that $u\in\W  kp\loc(M)$ is nonnegative and satisfies
\bea
-\Delta u +V u & \ge 0\\
\partial_\nu u+\mu u & \ge 0\ondm.
\eea
If $u(x)=0$ at some point $x\in M$, then $u$ vanishes identically.
\Proof 
From Sobolev embedding, we can assume without loss of generality
that the hypotheses of Proposition \ref{prop:linearprob} are satisfied
with $k=2$.  Suppose first that $x$ is an interior point of $M$.
Since $g$ is continuous and $V\in L^p\loc(M)$ with $p>n/2$, the weak 
Harnack inequality of \cite{trudinger-measurable} holds and we have for 
some radius $R$ sufficiently small and some exponent $q$ sufficiently 
large there exists a constant $C>0$ such that
\be
||u||_{L^q(B_{2R}(x)}\le C \inf_{B_R(x)} u = 0.
\ee
Hence $u$ vanishes in a neighbourhood of $x$, and a connectivity
argument implies $u$ is identically 0.

It remains to consider the case where $u$ vanishes at a point 
$x\in\partial M$.  Working in local coordinates about $x$
we can do our analysis on $B_1^+(0)\equiv B_1(0)\cap\Rn_+$,
where balls are now taken with respect to the flat background metric.
Let $b$ be a $\W 1p(B_1^+)$ vector field such that $\ip<b,\nu>=\mu$
on $D_1$, where $D_1$ is the flat portion of the boundary of $B_1^+$.
For example, since $\mu\in\W{1-\expfrac1p}p$ and $g\in\W2p\loc$,
we can take $b=\hat{\mu}\hat{\nu}$
where $\hat{\mu}$ is a $\W 1p$ extension of $\mu$ and $\hat{\nu}$
is a $\W 2p$ extension of $\nu$.  Integrating by
parts, we have for any nonnegative $\phi\in C\cpct^\infty(B_1^+\cup D_1)$
\be
\int_{B_1^+} \ip<\nabla \phi, b>u + u\phi\,\div b +\ip<b,\nabla u>\phi\dV =
\int_{D_1} \mu\,u \phi \dA.
\ee
Hence
\bean
\int_{B_1^+}\left<\nabla u,\nabla\phi\right>_g + R\,u\phi+{}\qquad\qquad
\qquad\qquad\nonumber\\
+ \ip<\nabla \phi, b>u + u\phi\,\div b +\ip<b,\nabla u>\phi \dV
& = 
\int_{B_1^+}-\phi\Delta u +R\,u\phi\dV+\hfill \nonumber\\
&\quad+ \int_{D_1} \partial_\nu u\phi +\mu\, \phi u \dA  \nonumber\\
& \ge 0,
\label{dirichlet-form}
\eean
since $u$ is a supersolution.

To reduce to the interior case, we now construct an elliptic equation on all 
of $B_1$. For any function or tensor $f$ defined on $B_1^+(0)$, let $\tilde{f}$
be the extension of $f$ to $B_1$ via its pushforward under reflection.  
It follows from \eqref{dirichlet-form} and a
change of variables argument that for any nonnegative
$\phi\in C\cpct^\infty(B_1)$,
\bel{dirichlet-form-ball}
\int_{B_1}\left<\nabla \tu,\nabla\phi\right>_{\tg} + \tilde{R}\,\tu\phi 
+\langle\nabla \phi, \tilde{b}\rangle_{\tg}\, \tilde{u} + \tu\phi\,\widetilde{\,\div b\,} 
+ \langle\tilde{b},\nabla \tilde{u}\rangle_{\tg}\phi \;\widetilde{dV} \ge 0.
\eel
Since $\tg\in\W1{2p}(B_1)$, $\tilde{R}\in L^p(B_1)$, $\tilde{b}\in L^{2p}(B_1)$,
and $\widetilde{\,\div b\,}\in L^p(B_1)$, we conclude from 
\eqref{dirichlet-form-ball} that $\tilde{u}$ is a weak $\W 1{2p}$ 
supersolution of 
an elliptic equation with coefficients having regularity considered 
by \cite{trudinger-measurable}.  Since $\tu\ge 0$ and $\tu(0)=0$, the
weak Harnack inequality again implies that $u$ vanishes in a
neighbourhood of $x$ and hence on all of $M$.
\EOProof

We now turn to the existence proof for the nonlinear problem
\eqref{fund-sys}.  We assume for simplicity that the nonlinearities
$F$ and $f$ have the form
\bea
F(x,y)&=\sum_{j=1}^l F_j(x)G_j(y) \cr
f(x,y)&=\sum_{j=1}^m f_j(x)g_j(y).
\eea

\prop\label{prop:subsup} Suppose
\bl
\li{1.} $(M,g)$ is asymptotically Euclidean of class $\W kp\srho$ with $k\ge2$, $p>n/k$,
and $\rho<0$,
\li{2.} $u_-,\;u_+\in\W k{p}\del$ with and $\delta\in(2-n,0)$ are a 
subsolution and a supersolution 
respectively of \eqref{fund-sys} such that $u_-\le u_+$,
\li{3.} each $F_j\in \W{k-2}p_{\delta-2}(M)$ and $f_j\in\W{k-1-\expfrac1p}p(\partial M)$,
\li{4.} each $G_j$ and $g_j$ are smooth on $I=[\inf(u_-),\sup(u_+)]$.  

\el
Then there exists a solution $u$ of \eqref{fund-sys} such 
that $u_-\le u \le u_+$.
\Proof 
\def\LV{\mathop{L_V}}
\def\FV{F_V}
\def\Bmu{\mathop{B_\mu}}
\def\fmu{f_\mu}
We first assume $k=2$ and $p>n/2$. Let
\bea
V(x)&=\sum_{j=1}^l \abs{F_j(x)}\abs{\min_I G_j'}\\
\mu(x) &= \sum_{j=1}^m \abs{f_j(x)}\abs{\min_I g_j'},
\eea
so that $V\in L^p_{\delta-2}(M)$, $\mu\in\W{1-\expfrac1p}p(\partial M)$,
and both are nonnegative.  Let
$\FV(x,y)=F(x,y)+V(x)y$ and $f_\mu(x,y)=f(x,y)+\mu(x)y$
so that $\FV$ and $\fmu$ are both non-decreasing in $y$.
Let $\LV=-\Lap+V$, and let 
$\Bmu=\evat{\left(\partial_\nu+\mu\right)}{\partial M}$.
From Proposition \ref{prop:linearprob} we have $(\LV,\Bmu)$
is an isomorphism acting on $\W2p_\delta$.

We construct a monotone decreasing sequence of functions 
$u_+=u_0\ge u_1 \ge u_2 \ge \cdots$ by letting
\bea
\LV u_{i+1} &= \FV(x,u_i) \\
\Bmu u_{i+1} &= \fmu(x,u_i).
\eea
The monotonicity of the sequence follows from the maximum principle 
and the monotonicity of $\FV(x,y)$ and $\fmu(x,y)$ in $y$.  The maximum
principle also implies $u_i\ge u_-$.

We claim the sequence $\{u_i\}_{i=1}^\infty$ is bounded in $\W 2p\del(M)$.
From Proposition \ref{prop:linearprob} we can estimate
\bel{subsupe1}
||u_{i+1}||_{\W 2p\del(M)}
\lesssim ||\FV(x,u_{i})||_{L^p_{\delta-2}(M)} 
            + ||\fmu(x,u_{i})||_{\W{1-\expfrac1p}p(\partial M)}.
\eel
Pick $q\in(p,\infty)$ such that
\be
{1\over p} - {1\over n } < {1\over q} < {1\over n },
\ee
which is possible since $p>n/2$.  Then 
\bel{subsupe2}
||\FV(x,u_{i})||_{L^p_{\delta-2}(M)} +
||\fmu(x,u_{i})||_{\W{1-\expfrac1p}p(\partial M)} 
\lesssim 1+||u_i||_{\W{1}q(U)}
\eel
for any fixed smooth bounded neighbourhood $U$ of $\partial M$.
From interpolation and Sobolev embedding we have for any $\epsilon>0$
\be
||u_i||_{\W{1}q(U)}\lesssim 
C(\epsilon)||u_i||_{\W{1}p(U)}+\epsilon ||u_i||_{\W{2}p(U)}.
\ee
A second application of interpolation then implies
\be
||u_i||_{\W{1}q(U)}\lesssim 
C(\epsilon)||u_i||_{L^p(U)}+\epsilon ||u_i||_{\W{2}p(U)}.
\ee
Hence
\bel{subsupe3}
||u_i||_{\W{1}q(U)}\lesssim 
C(\epsilon)||u_i||_{L^p\del(M)} + \epsilon ||u_i||_{\W 2p\del(M)}.
\eel
Since $u_-\le u_i \le u_+$, we have $||u_i||_{L^p\del(M)}$ is
uniformly bounded.
Combining \eqref{subsupe1}, \eqref{subsupe2} and \eqref{subsupe3}
we obtain, taking $\epsilon$ sufficiently small,
\be
||u_{i+1}||_{\W 2p\del} \le {1\over 2}||u_{i}||_{\W 2p\del}+C
\ee
Iterating this inequality we obtain a bound for all $i$
\be
||u_{i}||_{\W 2p\del} \le 
||u_{+}||_{\W 2p\del}+ 2C.
\ee
Hence some subsequence of $\{u_i\}_{i=1}^\infty$ (and by monotonicity, the 
whole sequence) converges weakly in $\W 2p\del$ to a limit $u_\infty$. 

It remains to see
$u_\infty$ is a solution of \eqref{fund-sys}.
Now $u_i$ converges strongly to $u_\infty$ in $\W{1}p_{\delta'}$ for any 
$\delta'>\delta$, and also converges uniformly on compact sets.
Hence for any $\phi\in C\cpct^\infty(M)$,
\bea
\int_M\left(\FV(x,u_i)-V(x)u_{i+1}\right)\phi\dV &\rightarrow 
\int_M F(x,u_\infty)\phi\dV\\
\int_{\partial M} \left(\fmu(x,u_i)-\mu(x)u_{i+1}\right)\phi\dA &\rightarrow 
\int_{\partial M} f(x, u_\infty)\phi\dA\\
\int_M \ip<\nabla u_{i+1}, \nabla \phi> \phi\dV
&\rightarrow
\int_M \ip<\nabla u_\infty, \nabla \phi>\dV.
\eea
So
\be
\int_M \ip<\nabla u_\infty, \nabla \phi>\dV =
\int_M F(x,u_\infty)\phi\dV + \int_{\partial M} f(x, u_\infty)\phi\dA,
\ee
and an application of integration by parts shows $u_\infty$ solves
the boundary value problem.  

The case $k>2$ now follows from a bootstrap using the previous result 
together with Proposition \ref{prop:linearprob}.
\EOProof

\section{Solving the Lichnerowicz Equation}
\seclabel{sec:lich}

We now prove the existence of solutions of the Lichnerowicz
equation
\beal{conformal-h-2}
-\Lap \phi + {1\over a}\left(R\phi -\abs{\sigma}^2\phi^{-3-2\dimc}\right) &= 0\\
\partial_\nu\phi+{1\over \dimc} 
h\phi-{2\over a}\sigma(\nu,\nu)\phi^{-1-\dimc} &= 0
\quad\hbox{on $\partial M$}.\\
\eeal
We first show that if $\lambda_{g'}>0$, then $(M,g')$ is conformally
equivalent to $(M,g)$, where $g$ satisfies $\lambda_g>0$, $R=0$ and $h<0$.
We then show that if $\lambda_g>0$, $R=0$, and  
$(n-1)h\le\sigma(\nu,\nu)\le0$, then there exist a sub/supersolution pair
for \eqref{conformal-h-2} and we apply Proposition~\ref{prop:subsup}
to obtain a solution.

The following proposition gives useful conditions equivalent to 
$\lambda_g>0$.  We define for compactly supported functions
\be
Q_g(f) = 
{
\int_M a\abs{\nabla f}^2 + R f^2\dV + \int_{\partial M} {a\over\dimc}h f^2\dA
\over
||f||_{L^{2^*}}^2.
},
\ee
Thus
\be
\lambda_g=\inf_{f\in C\cpct^{\infty}(M),f\not\equiv0} Q_g(f).
\ee
We also define 
$$\calP_\eta=\left((-\Delta+{\eta\over a}R),
\evat{(\partial_\nu+{\eta\over\dimc}h)}{\partial M}\right).
$$  
If $k\ge2$, $\delta<0$, and $k>n/p$, then $\calP_\eta$ is a continuous
map from $\W kp\del(M)$ to $\W {k-2}p_{\delta-2}(M)\times\W {k-1-\expfrac1p}p(\partial M)$.

\prop\label{prop:guage} Suppose $(M,g)$ is asymptotically Euclidean of 
class $\W k{p}\del$, $k\ge2$, $k>n/p$, and $2-n<\delta<0$.
Then the following conditions are equivalent:
\bl
\li{1.} There exists a conformal factor $\phi>0$ such that $1-\phi\in\W kp\del$
and such that $(M,\phi^{2\dimc}g)$ is scalar flat and has a minimal 
surface boundary.
\li{2.} $\lambda_g>0$.
\li{3.} $\calP_\eta$ is an isomorphism acting on $\W kp\del$ for each $\eta\in[0,1]$.

\el
\Proof
Suppose condition 1 holds.  Since $\lambda_g$ is a conformal invariant,
we can assume that $R=0$ and $h=0$.  We first make a conformal change
to a metric with positive scalar curvature.  Let ${\cal R}$ be any 
continuous positive element of $\W {k-2}p_{\delta-2}$
and  let $v$ be the unique solution given by 
Proposition~\ref{prop:linearprob} to
\beal{cf1}
-\Lap_{\tg} v = {1\over a}{\cal R} \\
\partial_{\nu} v = 0.
\eeal
The maximum principle implies $v\ge 0$, and hence $\phi=1+v>0$.
Letting $\hat{g}=\phi^{2\dimc} g$, it follows that $\hat R$
is positive and continuous, and that $\hat h=0$. From Sobolev 
embedding and a standard argument with cutoff functions, we have 
\be
||f||_{L^{2^*}}^2\lesssim ||\nabla f||_{L^2}^2 + ||f||_{L^2(K)}^2,
\ee
where $K$ is the compact core of $M$.  Since $\hat R$ is bounded below
on $K$ we find
\be
||f||_{L^{2^*}}^2\lesssim ||\nabla f||_{L^2}^2 + ||\hat{R}^{1/2} f||_{L^2}^2
\ee
and hence $\lambda_g>0$.

Now suppose condition 2 holds.  We claim that for 
$\eta\in[0,1]$, the kernel of $\calP_\eta$ is trivial.
Since $\calP_\eta$ has index 0, this implies $\calP_\eta$ is an isomorphism.

Suppose, to produce a contradiction, that $u$ is a nontrivial 
solution. From Lemma~\ref{kerneldecay} we have $u\in\W kp_{\delta'}$ for any 
$\delta'\in(2-n,0)$.  From Sobolev embedding we have $u\in\W 
12_{\delta'}(M)$.  Taking $\delta'< 1-n/2$, we can integrate by parts to obtain
\be
0=\int_{M} -au \Lap u + \eta R u^2\dV = 
\int_{M} a\abs{\nabla u}^2 +  \eta R u^2\dV +
\eta\int_{\partial M} {a\over\dimc} hu^2\dA.
\ee
Since $\int_{M} a\abs{\nabla u}^2\dV \ge 0$, and since $\eta\in[0,1]$
we see
\be
0 \ge \eta\left[\int_{M} a\abs{\nabla u}^2 +  R u^2\dV +
\int_{\partial M} {a\over\dimc} hu^2\dA\right].
\ee
Since $\W 12_{\delta'}(M)$ is continuously embedded in $L^{2^*}(M)$ 
we conclude that $Q_g(u)\le 0$.
Since $Q_g$ is continuous on $\W kp{\delta'}$, and since $C^\infty\cpct$
is dense in $\W kp_{\delta'}$ we find $\lambda_g\le 0$, a contradiction.

Now suppose condition 3 holds.  For each $\eta\in[0,1]$, let $v_\eta$
be the unique solution in $\W kp_\delta$ of
\be
\calP_\eta v_\eta=-\eta(R/a, h/\dimc).
\ee
Letting $\phi_\eta=1+v_\eta$ we see 
\beal{scalarmean2}
-\Lap \phi_\eta + {\eta\over a} R \phi_\eta & = 0\\
\partial_\nu \phi_\eta+{\eta\over \dimc} h\phi_\eta & = 0.\\
\eeal
To show $\phi_\eta >0$ for all $\eta\in[0,1]$, we follow \cite{lap-asympt-flat}.
Let $I=\{\eta\in[0,1]:\phi_\eta>0\}$.  Since $v_0=0$, we have
$I$ is nonempty.  Moreover, the set $\{v\in C^0\del : v > -1\}$ is open 
in $C^0\del$.  Since the map taking $\eta$ to 
$v_\eta\in C^0\del$ is continuous, $I$ is open. It suffices to show
that $I$ is closed.  Suppose $\eta_0\in\bar{I}$.  Then
$\phi_{\eta_0}\ge 0$.  Since $\phi_\eta$ solves \eqref{scalarmean2},
and since $\phi_{\eta_0}$ tends to 1 at infinity,
Lemma~\ref{p:strmaxpr} then implies $\phi_{\eta_0}>0$.  Hence $\eta_0\in I$
and $I$ is closed.

Letting $\phi=\phi_1$ we have shown $\phi>0$.  Since $\phi$ solves 
\eqref{scalarmean2} with $\eta=1$ it follows that $(M,\phi^{2\dimc}g)$ is 
scalar flat and has a minimal surface boundary.  Moreover, since
$v\in\W kp\del$ we see $(M,\phi^{2\dimc}g)$ is also of class 
$\W kp\del$.
\EOProof

\Remark  In the context of asymptotically Euclidean manifolds without boundary,
Theorem 2.1 of  \cite{lap-asympt-flat} claims that one can make a conformal 
change to a scalar flat asymptotically Euclidean metric if and only if
\bel{cbcond}
\int_M a\abs{\nabla f}^2 + R f^2\dV >0\quad\hbox{for all $f\in C\cpct^\infty(M)$
, $f\not\equiv0$}.
\eel
The proof of Theorem 2.1 in \cite{lap-asympt-flat} has a mistake, and the 
condition \eqref{cbcond} is too weak.  A similar claim and error 
appears in \cite{const-asympt-flat}.
We see from Proposition~\ref{prop:guage} that the correct condition is 
\be
\inf_{f\in C\cpct^\infty(M),f\not\equiv0}
{
\int_M a\abs{\nabla f}^2 + R f^2\dV 
\over
||f||_{L^{2^*}}^2} > 0.
\ee
\EORemark

\cor\label{cor:guage2}
Suppose $(M,g')$ is asymptotically Euclidean of class
$\W kp\del$, $k\ge2$, $k>n/p$, and $2-n<\delta<0$.  If
$\lambda_{g'}>0$, then there exists
a conformal factor $\phi>0$ such that $1-\phi\in\W kp\del$
and such that $(M,g)=(M,\phi^{2\dimc}g')$  is scalar flat, has negative
boundary mean curvature, and satisfies $\lambda_g>0$.
\Proof
Since $\lambda_{g'}>0$, from Proposition \ref{prop:guage} we can
assume without loss of generality that $(M,g')$ satisfies $R'=0$ and $h'=0$.

Let $v_\epsilon\in\W kp\del(M)$ be the unique solution of
\bea
-\Lap_{g'} v_\epsilon  & = 0\\
\partial_{\nu'} v_\epsilon & = -\epsilon.\\
\eea
Since $k>n/p$, $v_\epsilon$ depends continuously in $C^0_\delta$
on $\epsilon$. Since $v_0=0$, we have $v_\epsilon>-1$ for $\epsilon$
sufficiently small.  Fixing one such $\epsilon>0$ we have
$\phi=1+v_\epsilon>0$.  Letting $g=\phi^{2\dimc}g'$ we see that
$R=0$ and $h= -\epsilon \dimc \phi^{-\kappa-1} < 0$. Proposition~\ref{prop:guage}
shows that $\lambda_g$ is a conformal invariant and hence $\lambda_g>0$ also.
\EOProof

\thm \label{thm:main} Suppose $(M,g)$ is asymptotically Euclidean of 
class $\W kp\del$, $k\ge 2$, $k>n/p$, and $2-n<\delta<0$.  Suppose
also that $\lambda_g>0$, $R=0$, and $h\le0$.  If $\sigma\in\W
{k-1}{p}_{\delta-1}$ is a transverse traceless tensor 
on $M$ such that $(n-1)h\le\sigma(\nu,\nu)\le 0$ on $\partial M$, then
there exists a conformal factor $\phi$ solving \eqref{conformal-h-2}.
Moreover, setting $\hat g = \phi^{2\dimc} g$ and $\hat K = \phi^{-2} \sigma$,
we have that $(M,\hat g)$ is asymptotically Euclidean of class $\W kp\del$,
$\hat K\in\W{k-1}p_{\delta-1}$, $(M,\hat g,\hat K)$
solves the Einstein constraint equations with apparent horizon boundary 
condition, and $\partial M$ is a marginally trapped surface.
\Proof
Setting $\phi = 1+v$ and $\sigma'={2\over a}\sigma(\nu,\nu)$, 
the Lichnerowicz equation reduces to solving
\beal{lichreduced}
-\Lap v &= {1\over a}\abs{\sigma}^2(1+v)^{-3-2\dimc}\\
\partial_\nu v &= -{1\over\dimc}h(1+v)+\sigma'(1+v)^{-1-\dimc}\ondm
\eeal
with the constraint $v>-1$.  We solve this by means of 
Proposition \ref{prop:subsup}.
Since ${a\over 2\kappa}=n-1$, and since $(n-1)h\le\sigma(\nu,\nu)$,
we conclude $-{1\over\dimc}h+\sigma'\ge 0$.  Therefore $v_-=0$ 
is a subsolution  of \eqref{lichreduced}. To find a supersolution, we solve for
each $\eta\in[0,1]$
\beal{supeq}
-\Lap v_\eta&= {1\over a}\abs{\sigma}^2\\
\partial_\nu v_\eta +{\eta\over\kappa}hv_\eta&=  -{\eta\over\kappa}h.
\eeal
The solution exists since $\lambda_g>0$.  We claim moreover that
$\phi_\eta=1+v_\eta> 0$.   
Let $I=\{\eta\in[0,1]:\phi_\eta>0\}$.  Arguing as in Proposition 
\ref{prop:guage},
using the fact that $\abs\sigma^2\ge0$, we see $I$ is open and nonempty.
Suppose $\eta_0\in\bar{I}$.  Then $\phi_{\eta_0}\ge 0$.  Since 
$\phi_\eta$ satisfies
\bea
-\Lap \phi_\eta&\ge 0\\
\partial_\nu phi_\eta +{\eta\over\kappa}h\phi_\eta&=  0,
\eea
and since $\phi_{\eta_0}$ tends to 1 at infinity,
Lemma~\ref{p:strmaxpr} then implies $\phi_{\eta_0}>0$.  Hence $\eta_0\in I$
and $I$ is closed.  Let $v_+=v_1$.  We have proved $1+v_+> 0$.
But then, since
\beal{supeq2}
-\Lap v_+&= {1\over a}\abs{\sigma}^2\\
\partial_\nu v_+ &= -{1\over\dimc} h(1+v_+), \\
\eeal
and since $-h\ge0$, $1+v_+\ge 0$ and $\abs\sigma^2\ge0$, we conclude that
$v_+\ge 0$.  Now
\be
-\Lap v_+ = {1\over a}\abs{\sigma}^2\ge {1\over a}
\abs{\sigma}^2(1+v_+)^{-3-2\dimc},
\ee
and since $\sigma'\le 0$ we have
\be
\partial_\nu v_+ = -{1\over\dimc}h(1+v_+) \ge 
-{1\over\dimc}h(1+v_+)+\sigma'(1+v_+)^{-1-\dimc} \ondm.
\ee
So $v_+$ is a nonnegative supersolution of \eqref{lichreduced}.

Now $v_-$, $v_+$, $(M,g)$, and the right hand sides of \eqref{lichreduced}
all satisfy the hypotheses of Proposition \ref{prop:subsup}.
So there exists a nonnegative solution $v$ of \eqref{lichreduced} in 
$\W kp\del$. 
Letting $\hat g = \phi^{2\dimc} g$ and $\hat K=\phi^{-2}\sigma$,
it follows that  $(M,\hat g,\hat K)$ solves the constraint equations with 
apparent horizon boundary condition. 
To see that the boundary is marginally trapped, we note that 
\be
(n-1)\hat h=\hat K(\hat \nu,\hat \nu)=\phi^{-2-2\dimc}\sigma(\nu,\nu)\le0.
\ee 
Since $\hat\theta_-=\hat\theta_+ + 2(n-1)\hat h$, we conclude
$\hat\theta_-\le\hat\theta_+=0$, and $\partial M$ is marginally trapped.
\EOProof

The previous arguments and theorems can all be easily modified
for manifolds without boundary by omitting boundary conditions and
all references to the boundary of the manifold.  Hence we also have 
the following low regularity construction for manifolds without boundary.

\thm \label{thm:nobdy} Suppose $(M,g)$ is an asymptotically Euclidean 
manifold without boundary of 
class $\W kp\del$, $k\ge 2$, $k>n/p$, $2-n<\delta<0$, and suppose $\sigma\in\W
{k-1}{p}_{\delta-1}$.  If $(M,g)$ satisfies $\lambda_g>0$, 
then there exists a conformal factor $\phi$ 
solving \eqref{conformal-h-2}.  Moreover, $(M,\hat g)=(M,\phi^{2\dimc} g)$ 
is asymptotically Euclidean of class $\W kp\del$,
$\hat K=\phi^{-2}\sigma\in\W{k-1}p_{\delta-1}$, and $(M,\hat g,\hat K)$ 
solves the Einstein constraint equations.
\NoProof

\section{Constructing Suitable Transverse Traceless Tensors}
\seclabel{sec:vl}
We now prove the existence of a class of data satisfying the hypothesis 
of Theorem \ref{thm:main}.  
It is easy to construct asymptotically Euclidean manifolds satisfying
$\lambda_g>0$.  From the proof of Proposition~\ref{prop:guage},
it is clear that every manifold with $R\ge0$ and $h\ge0$ satisfies
$\lambda_g>0$. Let $(M,g,K)$ be an asymptotically Euclidean manifold
without boundary satisfying $R\ge0$ (for example any maximal
solution of the vacuum constraint equations).
Let $v$ be the Greens function for the operator $-a\Delta+R$ with pole at
$x\in M$, and let $\phi=1+v$. Since $R\ge0$, we have $v\ge 0$ and $\phi>0$.  
Setting $\tg=\phi^{2k}g$ we see that $(M-\{x\},\tg)$ satisfies $\tR\ge 0$. 
Moreover, if $B_r$ is a geodesic ball (with respect to $g$) about $x$ 
with radius $r$, one readily verifies 
from the asymptotic behaviour $\phi\approx r^{2-n}$ near $x$ that 
$\tilde{h}>0$  for $r$ sufficiently small.  So $(M-B_r,\tg)$ satisfies 
$\lambda_{\tg}>0$. In the same way, given an asymptotically Euclidean  
manifold with boundary satisfying $R\ge0$ and $h\ge0$ we can perform the
previous construction using  the Greens function for $-a\Delta+R$ 
corresponding to the boundary condition $\partial v +hv=0$ to add
another boundary component. Continuing iteratively, we can add as 
many boundary components as we like.

On the other hand, to provide a suitable tensor $\sigma$, we must do
more work.  The requirements on $\sigma$ are that it be trace-free,
divergence-free, and satisfy $(n-1)h\le\sigma(\nu,\nu)\le0$ on
$\partial M$.  We construct $\sigma$ using a boundary value problem
for the vector Laplacian.

Let $\ck$ denote the conformal Killing operator, so 
$\ck X = {1\over 2}\Lie X g-{1\over n}(\div X) g$.  Then the vector Laplacian
$\vl=\div\ck$ is an elliptic operator on $M$, and the Neumann boundary
operator $B$ corresponding to $\vl$ takes a vector field $X$ to the covector
field $\ck X(\nu,\cdot)$.  We propose to solve the boundary value problem
\beal{vlext}
\vl X & = 0 \\
B X & = \omega\ondm,
\eeal
where $\omega$ is a covector field over $\partial M$.
If we can do this, then letting $\sigma=\ck X$ it follows that $\sigma$
is trace and divergence free.  Moreover, $\sigma(\nu,\nu)= \omega(\nu)$ 
on $\partial M$, so taking $\omega$ such that $(n-1)h\le\omega(\nu)\le0$
ensure $(n-1)h\le\sigma(\nu,\nu)\le0$.

If $(M,g)$ is asymptotically Euclidean of class $\W kp\srho$
with $k\ge2$ and $k>n/p$, then $\vl$ and $B$ act continuously as 
maps from $\W kp\del(M)$ to $\W{k-2}p_{\delta-2}(M)$ and
$\W{k-1-\expfrac1p}p(\partial M)$ respectively.
We show \eqref{vlext} is well posed by proving 
$\PL=(\vl, B)$  is an isomorphism if $2-n<\delta<0$ and either 
$k>n/p+1$ or $k>n/p$ and 
$(M,g)$ has no nontrivial conformal Killing fields vanishing at infinity.

The method of proof is well established \cite{lap-weighted}
\cite{elliptic-systems-H}. The first step is to 
obtain a coercivity estimate (equation \eqref{coercive} of 
Proposition \ref{extcoercive} 
below)  that implies $\PL$ is semi-Fredholm.  The second step is to 
explicitly compute the index and dimension of the kernel of $\PL$,
which we do in Proposition \ref{vliso}.

In fact, for smooth metrics it follows from \cite{elliptic-sys-noncompact} that
$\PL$ is Fredholm.  So our principal concern is to show that the coercivity
estimate holds for low regularity metrics.

We start with an priori estimate for $\Leps$ on a compact 
manifold $K$.   We assume that the boundary of $K$ is partitioned into two 
pieces, $\partial K_1$ and $\partial K_2$, each the union of components of
$\partial K$ and either possibly empty.  We then have an estimate for
$\vl$ in terms of a Neumann condition on $K_1$ and a Dirichlet 
condition on $K_2$.

\prop\label{compactbound} Suppose $(K,g)$ is a compact manifold with boundary
of class $\W kp$ with $k>n/p$ and $k\ge 2$. If $X\in\W kp(K)$, then 
\beal{cpctest}
||X||_{\W kp(K)}&\lesssim ||\Leps X||_{\W{k-2}p(K)}+
||\Beps X||_{\W{k-1-\expfrac1p}p(\partial K_1)}+\\
&\qquad+||X||_{\W{k-\expfrac1p}p(\partial K_2)}+||X||_{L^p(K)}.
\eeal
\Proof
The estimate follows from interior and boundary
estimates together with a partition of unity argument.  We
explicitly prove the local estimate near $\partial K_1$, the 
other estimates being similar and easier.
In local coordinates near some fixed $x\in \partial K$  
we may assume the boundary is flat, $x=0$,
and  $g_{ij}(0)=\delta_{ij}$.  In these coordinates, 
\be
\vl X_j = \sum_{i,j,\abs{\alpha}\le 2} c\jia\partial_\alpha X_i
\ee
where $c\jia\in\W{k-2+\abs\alpha}p\loc$.

Suppose first that $X\in\W kp(B_r^+)$ 
with support contained in $B_{r/2}^+$, where $r$ is a small 
number to be specified later.
Let $\Lflat$ and $\Bflat$ denote the constant coefficient operators
given by the principal symbols of $\Leps$ and $\Beps$ at $0$;
these are in fact the vector Laplacian and Neumann boundary
operator computed with respect to the Euclidean metric on the
half space.  It is easy to verify that these satisfy the 
so-called Lopatinski-Shapiro or covering conditions of \cite{adn}
and hence we have
\bel{adnbound}
||X||_{\W kp(\hBr)}\le ||\Lflat X||_{\W{k-2}p(\hBr)} +
||\Bflat X||_{\W{k-1-\expfrac1p}p(D_r)}+||X||_{\W{k-2}p(\hBr)},
\eel
where $D_r$ is the flat portion of $\partial B_r^+$.
In local coordinates 
\be
(\Lflat X)_j= (\Leps X)_j 
+\sum_{\abs{\alpha}= 2} (\overline{c}\jia-c\jia)\partial_\alpha X_i
+\sum_{\abs{\alpha}< 2} c\jiapa X_i.
\ee
We wish to estimate each of these terms in $\W{k-2}p(\hBr)$, and
we must proceed carefully since a naive 
application of the multiplication rule from Lemma \ref{basic} will introduce 
unwanted large terms involving $||X||_{W^{2,p}}$.  Consider a term of 
the form $c\jiapa X_i$ with $\abs\alpha<2$.  Pick $q\in(p,\infty)$ 
such that 
\be
{1\over p} - {1\over n} < {1 \over q } < {k-1\over n}.
\ee
Then applying Lemma \ref{basic} we have
\be
||c\jiapa X_i||_{\W {k-2}p(\hBr)} \lesssim ||X_i||_{\W {k-1} q(\hBr)}
\ee
and, arguing as in Proposition \ref{prop:subsup},
we obtain from interpolation and Sobolev embedding
\be
||c\jia\partial_\alpha X_i||_{\W {k-2}p(\hBr)} \lesssim 
C(\epsilon)||X||_{L^p(\hBr)}+ \epsilon ||X||_{\W kp(\hBr)}.
\ee
The terms involving $(\overline{c}\jia-c\jia)\partial_\alpha X_i$ are
estimated similarly, and we have
\be
(\overline{c}\jia-c\jia)\partial_\alpha X_i
\lesssim C(\epsilon)||X||_{L^p(\hBr)} +
\left(||\overline{c}\jia-c\jia||_{L^\infty}+\epsilon\right)||X_i||_{\W kp(\hBr)}.
\ee
Since $||\overline{c}\jia-c\jia||_{L^\infty}$ can be made as small
as we please by taking $r$ small enough we conclude
\bel{lflatbound}
||\Lflat X||_{L^p(\hBr)} \lesssim 
||\Leps X||_{L^p(\hBr)}+C(\epsilon)||X||_{L^p(\hBr)}+\epsilon||X||_{\W{2p}(\hBr)}.
\eel
A similar argument with the boundary operator leads us to
\beal{bflatbound}
||\Bflat X||_{\W{k-1-\expfrac1p}p(D_r)} &\lesssim 
||(\Beps X)_j||_{\W{k-1-\expfrac1p}p(D_r)} +\\
&\qquad C(\epsilon)||X||_{\W {k-1}p(\hBr)}+
\epsilon||X||_{\W kp(\hBr)}.
\eeal
Combining \eqref{adnbound}, \eqref{lflatbound} and \eqref{bflatbound},
taking $\epsilon$ sufficiently small, we conclude
\be
||X||_{\W kp(\hBr)}\lesssim ||\Leps X||_{\W{k-2}p(\hBr)}+
||\Beps X||_{\W{k-1-\expfrac1p}p(D_r)}+||X||_{\W {k-1}p(\hBr)}.
\ee

The estimate \eqref{cpctest} for $X$ with arbitrary support is 
now achieved in a standard way with cutoff functions and a partition 
of unity argument. 
\EOProof

With the a priori estimate for compact manifolds in hand, the following 
coercivity result is now standard
\cite{global-asympt-simple}
\cite{elliptic-systems-H}\cite{mass-asympt-flat}.
We omit the proof for the sake of brevity.

\prop\label{extcoercive}  
Suppose $(M,g)$ is asymptotically Euclidean of class $\W kp\srho$
with $k>n/p$ and $k\ge 2$.  Then if $2-n<\delta<0$, 
$\delta'\in\Reals$, and $X\in\W kp\del(M)$ we have
\bel{coercive}
||X||_{\W kp\del(M)}\lesssim ||\Leps X||_{\W {k-2}p_{\delta-2}(M)}
+||\Beps X||_{\W{k-1-\expfrac1p}p(\partial M)}+||X||_{L^p_{\delta'}(M)},
\eel
noting that the inequality is trivial if $||X||_{L^p_{\delta'}(M)}=\infty$.
\NoProof

Let $\P kp\delta$ denote $\PL$ acting
as a map from $\W kp\del(M)$ to 
$
\W{k-2}p_{\delta-2}(M)\times\W{k-1-\expfrac1p}p(\partial M)$.

From estimate \eqref{coercive} it follows immediately \cite{schechter-fa}
that $\P kp\delta$ is semi-Fredholm under the assumptions
on $k$, $p$, and $\delta$ of Proposition \ref{extcoercive}.  
We now show $\P kp\delta$ is Fredholm with index 0.

\prop\label{vliso} Suppose $(M,g)$ is asymptotically Euclidean of class
$\W kp\srho$ with $k>n/p$, $k\ge 2$, and suppose 
$2-n<\delta<0$. Then $\P kp\del$ is Fredholm of index 0. Moreover, 
it is an isomorphism if and only if $(M,g)$ possesses no nontrivial 
conformal Killing fields in $\W kp\del(M)$.
\Proof
We first suppose $(M,g)$ is asymptotically flat of class $C^\infty\srho$; 
the desired result for rough metrics will follow from an index theory
argument.

It is enough to prove that $\P22\delta$ is invertible.  Indeed, from
elliptic regularity we know that the the kernels of $\P kp\delta$
and $\P22\delta$ agree.  And if $\P22\delta$ is surjective,
then the image of $\P22\delta$ contains
$C^\infty\cpct(M)\times C^\infty(\partial M)$. Again from elliptic
regularity we have that the image of $\P kp\delta$ also 
contains $C^\infty\cpct(M)\times C^\infty(\partial M)$ and since
the image of $\P kp\delta$ is closed, $\P kp\delta$ is surjective.

We restrict our attention now to $\calP=\P 22\delta$. To show $\calP$ 
is injective, we prove that any element of the kernel of $\calP$ is a 
conformal killing field.
Since $g$ is smooth, it follows from \cite{boost-problem} (see also
Section \ref{sec:nock}) that there are no nontrivial conformal Killing fields
in $\W 22\del$, and hence the kernel of $\calP$ is trivial.
Suppose $u\in\ker\calP$.We would like to integrate by parts to deduce that
\be
0 = -\int_M <\Leps u,u> \;dV = \int_M  <\ck u,\ck u>\;dV
\ee
and hence $u$ is a conformal Killing field.
This computation is only valid if $\delta\le {1-{n\over2}}$.
It was shown in \cite{boost-problem} (in the case of manifolds 
without boundary) that if $u\in\ker\calP$, then 
$u\in\W 22_{\delta'}$ for every $\delta'\in(2-n,0)$.  
Their proof also works if $\calP u$ is only compactly supported, and
a simple hole filling argument then implies the same result holds
for manifolds with boundary.  So $\calP$ is injective.

To show $\calP$ is surjective, it is enough to show
that the adjoint $\calP^*$ is injective.  The dual space of
$L^2_{\delta-2}(M)\times H^{1/2}(\partial M)$ is
$L^{2}_{2-n-\delta}(M)\times H^{-1/2}(\partial M)$.
From elliptic regularity \cite{hormander-three}
and rescaled interior estimates we know that 
if $\calP^*(f,h)=0$, then in fact $f$ and $h$ are smooth and
$f\in H^2_{2-n-\delta}(M)$.
Now if $\phi$ is smooth and compactly supported in each end of $M$
we have from integrating by parts
\bean
0 &= \left<\calP(f,h),\phi\right>\nonumber\\
&= \int_{M} <\Leps f,\phi> \,dV + 
\int_{\partial M} \ck\phi( \nu, f+h )- \ck f( \nu, \phi )\,dA.
\label{ibp}
\eean
We obtain immediately $\Leps f = 0$ in $M$.  
Moreover, one can readily show that if $\omega$ is a smooth 1-form
on $\partial M$ and $\psi$ is a smooth function on $\partial M$,
then there exists a $\phi\in C^\infty\cpct$ such 
that $\phi=\psi$ and $\ck\phi=\omega$ on $\partial M$.
It follows that $B f=0$ and $h=0$.
Since $\Leps f=0$ and $B f=0$, we have $f=0$. Hence $\calP^*$ is 
injective and therefore $\calP$ is an isomorphism.

We return to the case where $g$ is not smooth but only in $\W kp\srho(M)$ 
with $k>n/p$ and $k\ge2$.  To show $\P kp\delta$ is Fredholm of index 0,
it is enough to show its index is 0.  Since $g$ can be approximated
with smooth metrics $g_k$, and since each $\P kp{\delta,g_k}$ has index 0,
so does the limit $\P kp\delta$.  To show that the kernel of $\P kp\delta$
consists of conformal Killing fields, we integrate by parts again using
the fact that elements in the kernel decay sufficiently fast at infinity.
\EOProof

Proposition \ref{vliso} reduces the question of whether or not $\PL$ is an isomorphism 
to the existence of nontrivial conformal Killing fields $X$ vanishing
at infinity with $\ck X(\nu,\cdot)=0$ on $\partial M$.  In \cite{boost-problem}
it was proved that if $(M,g)$ is asymptotically Euclidean of class $\W kp\srho$
with $k> n/p+3$ that there are no nontrivial conformal Killing fields
in $\W kp\del$ for $\delta<0$.  Hence the boundary value problem
\eqref{vlext} is well posed if the metric has this high degree of regularity.
It has been subsequently claimed \cite{const-asympt-flat} that no
nontrivial conformal Killing fields exist with the metric as irregular as
$k>n/p+2$, but no proof exists in the literature.  The following section
contains a proof that there are no such vector fields for metrics as
irregular as $k>n/p+1$ and thereby establishes
\thm\label{vlwellposed}
Suppose $(M,g)$ is asymptotically Euclidean of class $\W kp\srho$
with $k\ge2$ and suppose $2-n<\delta<0$.  If either $k>n/p+1$ or 
$k>n/p$ and $(M,g)$ has no nontrivial conformal Killing fields in
$\W kp\del(M)$, then there exists a unique solution 
$X\in\W kp\del(M)$ of \eqref{vlext}.
\NoProof

\section{Non-Existence of Conformal Killing Fields Vanishing at Infinity}
\seclabel{sec:nock}

We break the problem of showing conformal Killing fields vanishing at
infinity are trivial into two pieces as was done in the high
regularity proof of \cite{boost-problem}. We first show that if a
conformal Killing field exists, it must be identically zero in a
neighbourhood of each end.  We then show that the zero set can
be extended to encompass the whole manifold.  For both parts of the
argument, we use a blowup method to construct a conformal
Killing field on a subset of $\Rn$ and analyze properties of the
resulting limit field.

In this section we use the notation $e_i$ for a standard
basis element of $\Rn$ or $\Reals^{n+1}$.  Recall \cite{conformal-geometry} 
that a basis for the conformal Killing fields on $\Rn$ with the Euclidean 
metric is comprised of the generators of the translations $\{e_i\}_{i=1}^n$, 
the rotations 
$\{x_i e_j - x_j e_i\}_{1\le i<j\le n}$ and the spherical dilations
$\{D_{e_k}\}_{k=1}^{n+1}$.  
Our notation for the dilations is as follows.
Any constant vector $V$ in $R^{n+1}$ gives rise to a function 
$x\mapsto<V,x>$ on the sphere.  The field $D_V$ is the pushforward under
stereographic projection of the gradient of this function.  We note
that $D_{V_1}+D_{V_2}=D_{V_1+V_2}$.  Moreover, in local coordinates we have
\bel{dilation}
D_{e_1} = ( {1\over2}\left(1-x_1^2+x_2^2+\cdots+x_n^2\right), -x_1 x_2, 
\cdots, -x_1 x_n). 
\eel
The dilations $D_{e_j}$ for $1<j\le n$ have similar coordinate expressions being
the pushforward of $D_{e_1}$ under a rotation. Finally, $D_{e_{n+1}}$ has
the coordinate expression 
\be
D_{e_{n+1}} = (x_1,\cdots,x_n).
\ee

The following two lemmas provide properties of conformal Killing fields
on subsets $\Rn$.  The first one can also be deduced from the analysis
in \cite{boost-problem}, but we include it here for completeness.

\lem \label{cfflatext} Suppose $X$ is a conformal Killing field on
the external domain $E_1$ with the Euclidean metric.  Suppose moreover 
that $X\in L^p\del(E_1)$ for some $p\ge1$ and $\delta<0$.  
Then $X$ vanishes identically.
\Proof
The basis of conformal Killing fields on $\Rn$ restricts to a
basis of conformal Killing fields on $E_1$. If $X$ is a sum of such 
vectors, its coefficients are polynomials. But no non-zero polynomial is 
in $L^p\del(E_1)$, since $\delta<0$.  Hence $X=0$.
\EOProof

\lem\label{ckflatclass} Suppose $X$ is a nontrivial conformal 
Killing field on $B_1(\Rn)$ with the Euclidean metric that satisfies
$X(0)=0$ and $\nabla X(0)= 0$. Then $X(x)\neq0$ if $x\neq 0$.
\Proof
By a conformal change of metric, it is enough to show the same
is true on $\Rn$. A routine computation shows that the subspace 
of conformal Killing vector fields in $\Rn$ that vanish at 0 is 
spanned by the rotations, the dilation $D_{e_{n+1}}$, and the vectors
\bel{shifteddilations}
2D_{e_i}-e_i\qquad i=1\cdots n.
\eel
The coefficients of the vectors \eqref{shifteddilations}
are homogeneous polynomials of degree 2 and hence their
derivatives at the origin vanish.  
On the other hand, a vector in the span of $D_{e_{n+1}}$ and the 
rotations has linear coefficients and vanishes identically 
if its gradient at the origin vanishes.
So if $X(0)=0$ and $\nabla X(0)=0$,
then $X$ is in the span of of the vectors \eqref{shifteddilations}.
We claim a nontrivial such vector vanishes only at the origin.  Indeed,
\be
\sum_{j=1}^n v^j \left( 2D_{e_j}-e_j\right) = 2D_V - V
\ee
where $V=\sum_{j=1}^n v^j e_j$. If $V$ is not zero, then by performing a 
rotation and constant scaling we can push $2D_V-V$ forward to 
$2D_{e_1}-e_1$. From the explicit expression \eqref{dilation} we observe that 
$2D_{e_1}-e_1$ vanishes only at 0.
\EOProof

We say a vector field $X$ 
\define{vanishes in a neighbourhood of infinity}
if for each end there exists a radius $R$ such that in end coordinates
$X\equiv 0$ in the exterior region $E_R$.  The following lemma 
shows under weak hypotheses on the metric that a conformal Killing field 
vanishing at infinity also vanishes in a neighbourhood of infinity.

\lem\label{ckends} Suppose $(M,g)$ is asymptotically Euclidean of 
class $\W 2p\srho$ with $p>n/2$.  Suppose $X$ is a conformal 
Killing field in $\W 2p\del$ with $\delta<0$.  Then $X$ vanishes in
a neighbourhood of infinity.
\Proof
We work in end coordinates and construct a sequence 
of metrics $\{g_k\}_{k=1}^{\infty}$ on the exterior region $E_1$ by 
letting $g_k(x) = g(2^k x)$.  Since 
$||g_k-\gflat||_{\W2p\srho(E_1)}
\lesssim 2^{\rho k}||g-\gflat||_{\W2p\srho(E_{2^k})}$
we conclude $||g_k-\gflat||_{\W2p\srho(E_1)}\rightarrow 0$.
It follows that the associated maps $\vl^k$ and $\ck^k$ acting
on $\W2p\del(E_1)$ converge as operators to $\vlflat$ and $\ckflat$.

Suppose, to produce a contradiction, that $X$ is not identically 0
outside any exterior region $E_R$.  Let  $\hat{X}_k$ be the vector
field on $E_1$ given by $\hat{X}_k(x)=X(2^k x)$ and let 
$X_k=\hat{X}_k/||\hat{X}_k||_{\W 2p\del}$.  Fix $\delta'$ with
$\delta'\in(\delta,0)$.  Then from the $\W2p\del$ boundedness 
of the sequence $\{X_k\}_{k=1}^\infty$ it follows 
(after reducing to a subsequence) that
the vectors $X_k$ converge strongly in $\W1p_{\delta'}$ to some $X_0$.
We can assume without loss of generality that $\delta> 2-n$ and hence
we can apply 
Proposition \ref{extcoercive} to the Euclidean metric to obtain
\bea
|| X_{k_1}-X_{k_2}||_{\W2p\del} & \lesssim ||\vlflat -\vl^{k_1}||_{\W2p\del} +
||\vlflat -\vl^{k_2}||_{\W2p\del} +\\
&\qquad +||\ckflat-\ck^{k_1}||_{\W2p\del}+
||\ckflat-\ck^{k_2}||_{\W2p\del}+||X_{k_1}-X_{k_2}||_{L^p_{\delta'}}.
\eea
Here we have used the $\W2p\del$ boundedness of the sequence 
$\{X_k\}_{k=1}^\infty$ together with the facts $\vl^k X_k=0$ and $\ck^k X_k=0$.
We conclude $\{X_k\}_{k=1}^\infty$ is Cauchy in $\W2p\del(E_1)$ 
and hence $X_k\converges{\W2p\del}X_0$.  Moreover, since 
$\ck^k\converges{\W2p\del}\ck$ and since $\ck^k X_k=0$
we have $X_0$ is a conformal Killing field for $\gflat$ in $\W2p\del(E_1)$.
From Lemma \ref{cfflatext} it follows that $X_0=0$.  Hence $X_k$ converges 
in $\W2p\del$ to 0, which contradicts $||X_k||_{\W2p\del}=1$ for each $k$.
\EOProof

\thm\label{nock} 
Suppose $(M,g)$ is asymptotically Euclidean of class $\W 2p\srho$
with $p>n$.  Then there exist no nontrivial conformal 
Killing fields in $\W kp\del$ for any $\delta<0$.
\Proof
From Lemma \ref{ckends} we know that if $X\in\W 2p\del$ is a conformal Killing
field then it vanishes on an open set.  We claim $X^{-1}(0)=M$.

If the claim is not true, then there exists a point $x_0$ in the interior 
of $M$ and on the boundary of the interior of $X^{-1}(0)$.  Working 
in local coordinates near $x$, we reduce to the situation where $g$ 
is a metric on the unit ball, $g_{ij}(0)=\delta_{ij}$, $X$ is a 
conformal Killing field for $g$ on $B_1$, and the origin is on the boundary
of the interior of $X^{-1}(0)$.  Since $p>n$ it follows that $X$ is 
in $C^1(B_r)$ and hence $X(0)=0$ and $\nabla X(0)=0$.

Let $\{r_k\}_{k=1}^\infty$ be a sequence of radii $r_k$ tending down to zero such that
$X(x)=0$ for some $x$ with $|x|=r_k/2$.  We construct
a sequence of metrics $\{g_k\}_{k=1}^\infty$ on the unit ball by taking 
$g_k(x) = g(x/r_k)$.  Evidently, $g_k\converges{\W2p} \overline{g}$, 
and it follows that the associated maps 
$\vl^k$ and $\ck^k$ converge to $\vlflat$ and $\ckflat$ as operators
on $\W 2p\del(B_1)$.

We construct vector fields $X_k$ on $B_1$ by setting 
$\hat{X}_k(x)=X(x/r_k)$ and letting $X_k=\hat{X}_k/||\hat{X}_k||_{\W 2p}$
(since $X$ is not identically 0 on $B_{r_k}$ the normalization is possible).  
By our choice of radii, there exists a point $x_k$ with $|x_k|=1/2$ such 
that $X_k(x_k)=0$.  From the $\W2p$ boundedness of the sequence $\{X_k\}_{k=1}^\infty$,
it follows (after taking a subsequence) that $X_k$ converges 
strongly in $\W1p$ to some $X_0\in\W2p$.

Arguing as in Lemma \ref{ckends}, replacing the use of Proposition 
\ref{extcoercive} with Proposition \ref{compactbound}, we conclude $X_0$ 
is a conformal killing field
for $\gflat$ and $X_k$ converges in $\W2p$ to $X_0$.
From the resulting $C^1(B_1)$ convergence of $X_k$ to $X_0$ 
we know $X_0(0)=0$ and $\nabla X_0(0)=0$.  The collection of points 
$x_k$ where $X_k$ vanishes has a cluster point $x$ with $|x|=1/2$,
and hence $X_0(x)=0$.  But Lemma \ref{ckflatclass} then implies $X_0=0$, which
is impossible since $||X_k||_{\W2p}=1$ for each $k$.  Hence 
$X=0$ identically.
\EOProof

\Remark Even though Theorem \ref{nock}
requires $\W2{n+\epsilon}$ regularity, we know from Lemma \ref{ckends}
that a conformal Killing field that vanishes at infinity also vanishes
in a neighbourhood of infinity, even if the metric
only has $\W2{\expfrac n2+\epsilon}$ regularity. Since
Theorem \ref{nock} is a unique continuation argument, it is perhaps
not surprising
that it requires $g\in\W2{n+\epsilon}\loc$, as this is the minimal
regularity that guarantees the principal coefficients of $\vl$
are Lipschitz continuous.  It would be interesting to determine
if Theorem \ref{nock} also holds in the low regularity case $\W2{\expfrac n2+\epsilon}$.
\EORemark

\Acknowledge
I would like to thank D. Pollack, J. Isenberg, and S. Dain for helpful 
discussions and advice. I would also like to thank an anonymous 
referee for suggestions that improved the paper's style. This 
research was partially supported by NSF grant DMS-0305048.
\EOAcknowledge

\references
\ifjournal
\bibliographystyle{journal}%
\else
\advance\baselineskip by -2 pt
\bibliographystyle{amsalpha-abbrv}%
\fi
\bibliography{mrabbrev,ahconstraints}
\endmatter